\documentclass[]{revtex4}% Physical Review B

\usepackage{graphicx}
\usepackage{amssymb, amsfonts}
\usepackage{subfig}
\usepackage{color}% bold math
\usepackage{bm}% bold math
\usepackage{url}
\usepackage{amsthm}
\usepackage[fleqn]{amsmath}
\usepackage{mathtools}

\begin{document}
\urlstyle{same}
\bibliographystyle{apsrev}

\title[Covid-19 in California]{Investigating Causal links from Observed Features in the first COVID-19 Waves in California} 

\author{Sarah Good$^1$,  Anthony~O'Hare$^1$}
\affiliation{$^1$ Computing Science and Mathematics, Faculty of Natural Sciences, University of Stirling, Stirling, UK}
 \date{\today}

\begin{abstract}
Determining who is at risk from a disease is important in order to protect vulnerable subpopulations during an outbreak. We are currently in a SARS-COV-2 (commonly referred to as COVID-19) pandemic which has had a massive impact across the world, with some communities and individuals seen to have a higher risk of severe outcomes and death from the disease compared to others.  
These risks are compounded for people of lower socioeconomic status, those who have limited access to health care, higher rates of chronic diseases, such as hypertension, diabetes (type-2), obesity, likely due to the chronic stress of these types of living conditions. Essential workers are also at a higher risk of COVID-19 due to having higher rates of exposure due to the nature of their work.

In this study we determine the important features of the pandemic in California in terms of cumulative cases and deaths per 100,000 of population up to the date of 5 July, 2021 (the date of analysis) using Pearson correlation coefficients between  population demographic features and cumulative cases and deaths. The most highly correlated features, based on the absolute value of their Pearson Correlation Coefficients in relation to cases or deaths per 100,000, were used to create regression models in two ways: using the top 5 features and using the top 20 features filtered out to limit interactions between features. These models were used to determine a) the most significant features out of these subsets and b) features that approximate different potential forces on COVID-19 cases and deaths (especially in the case of the latter set). Additionally, co-correlations, defined as demographic features not within a given input feature set for the regression models but which are strongly correlated with the features included within, were calculated for all features. 
%From determining the most highly correlated features to COVID-19 outcomes and which sets of features correlate with each other, the underlying factors behind these outcomes can be hypothesised.

The five features which had the highest correlations to cumulative cases per 100,000 were found to be the following: Overcrowding (\% of households), Average Household Size, Hispanic ethnicity (\% of population), Ages 0-19 (\% population), education level of 9th to 12th with no high school diploma (\% of population older than 25 years), and incidence rates of Long-term Diabetes Complications (per 100,000 population). For cumulative deaths per 100,000, the feature set was similar except Overcrowding (\% of households) replaced Long-term Diabetes Complications. The feature set for uncorrelated features was the same for both cases and deaths. This set was comprised of Overcrowding (\% of households), Wholesale trade (\% of workforce employed in), `Transportation, warehousing, and utilities' (\% of workforce employed in), and `Graduate or professional degree' (\% of population older than 25 years).
%All linear regression models were found to be significant overall. Model 1 (for cases per 100,000) only significant feature for predicting cases per 100,000 is Hispanic ethnicity in predicting cumulative cases per 100,000. The significant features in Model 2 were Transportation, warehousing, and utilities and `Graduate or professional degree'. In Model 3 (for deaths per 100,000), the significant features were Hispanic ethnicity and Average Family Size, whereas in Model 4, these features were `Transportation and warehousing, and utilities' and Overcrowding.
%Co-correlations for all features within Models 1 and 3 include Ages 60-69 (\% population), Ages 80+ (\% population), `Commuting -  Worked from home' (\% workers), health insurance status (\% non-institutionalised civilians), and `Less than 9th grade education' (\% population $>$ 25 years). For Models 2 and 4, the features co-correlated with at least 3 of the input features include Long-term Diabetes Complications (incidence per 100,000), 9th to 12th grade, no diploma (\% population $>$ 25 years), and Ages 0-19 (\% population).

\end{abstract}

%\pacs{}


\smallskip

\maketitle

\section{Introduction}

%The recent COVID-19 pandemic has highlighted the need for efficient and effective analysis of infection data to identify high risk individuals and for constructing effective control measures. Features are often identified from calculating correlations population level demographics and those infected with a disease. This simple high level analysis cannot capture the deeper, more subtle, confounding factors that explain transmission and who may be at risk (``correlation does not mean causation"). This paper will perform analysis on the first wave of COVID-19 to extract important features and will identify and reconcile what looks like inconsistencies in the results.

Coronavirus 2019 (COVID-19) is the respiratory illness caused by the virus SARS-COV-2. It was discovered in Wuhan China in late 2019 and usually causes mild respiratory symptoms and flu-like symptoms, including (but not limited to) as fever, dry cough, and fatigue.
It is mainly spread by air when infected individuals exhale, such as through breathing, talking, coughing, and sneezing \cite{Wang_2021,greenhalgh_ten_2021}. There is a range of effects of the disease with some suffering little to no symptoms and others, including those who are older, those who have preexisting health conditions or both, are more likely to suffer more severe illness, become hospitalised, and die should they become infected \cite{halvatsiotis_demographic_2020}.
The impact of COVID-19 has been immense worldwide. 
By July 1, 2021, World Health Organization (WHO) estimated that over 181 million cases and almost 4 million COVID-19 related deaths had been confirmed since the beginning of the pandemic \cite{WHO_dashboard}. 
On the same date, the Centre for Disease Control and Prevention (CDC) in the United States reported over 33 million cases and over 600 COVID-19 related deaths had been recorded\cite{CDC_tracker}. 
It is expected that that COVID-19 will reduce the life expectancy of Americans by over a year, with the effect being three to four times as large for Black and Latino populations compared to Whites \cite{andrasfay_reductions_2021}.
These risks are compounded for people of lower socioeconomic status and those who have limited access to health care. 
This is generally due to greater barriers to obtaining care \cite{liang_health_2019,cole_health_2018,mcmorrow_determinants_2014} as well as the higher rates of chronic diseases, such as hypertension, diabetes (type-2), obesity, and other illnesses in these communities, likely due to the chronic stress of these types of living conditions \cite{marmot_health_1991,algren_associations_2018}. 

In the United States in particular, health care access is limited by insurance status in addition to the availability of local hospitals and other health services in general. Those living in poverty are more likely to lack health insurance. For example, 11.0\% of all Californians between the ages 19 to 64 are uninsured compared to the 17.8\% of Californians in the same age group who live below the Federal poverty line \cite{KFF_HealthInsurance_poverty,KFF_HealthInsurance}. In a nationwide poll conducted by the Kaiser Family Foundation, 73.7\% of uninsured Americans cite the cost of insurance as the reason for their uninsured status, even after the implementation of the Affordable Care Act (ACA) \cite{KFF_Tolbert}. Uninsured Americans have limited access to health-care, especially preventative care services and treating chronic conditions compared to those who have health insurance \cite{liang_health_2019,cole_health_2018,mcmorrow_determinants_2014}.

This is especially worrying since people in poverty are more vulnerable to severe outcomes of COVID-19. Lower social status is associated with poorer health outcomes than those of higher status, including greater levels of stress and greater risk of chronic conditions such as obesity, hypertension, and diabetes \cite{marmot_health_1991,algren_associations_2018}]. Moreover, going through periods of unemployment (which was more common during the pandemic than before) increases the risk of metabolic syndrome \cite{Brunner2009SocialOS}. Because such conditions are known risk factors for severe disease, low income populations are more susceptible to severe disease and death when they are infected \cite{CDC_Covid19}.
Inadequate housing is also a significant risk factor for poorer health outcomes in general and COVID-19. 
It has been well established that overcrowded and substandard housing conditions are correlated with higher rates of diseases, including respiratory diseases like tuberculosis, and worse mental health outcomes \cite{stein_study_1950,thomson_health_2001,sharfstein_is_2001,pevalin_impact_2017}. Poor housing conditions have now also been linked to higher rates of COVID-19 case rates and mortality \cite{ahmad_association_2020}. 
Furthermore, overcrowded households are more vulnerable to COVID-19 as the virus is spread more easily indoors and when the infected person is in proximity to others \cite{wang_airborne_2021,greenhalgh_ten_2021}. All of these factors raise the risk of COVID-19 transmission and severe disease for populations in poor living conditions, as COVID-19 is also a respiratory illness.

Education, both in terms of work opportunities and health literacy, is also a determinant of health over one's lifetime. 
Lower levels of education limit one's opportunities in achieving higher employment and are also associated with lower levels of health literacy, or the skills needed to understand and assess health-related information \cite{Easton2010, Berkman2011}. 
Low health literacy has also been connected to increased risk of hospital admission for preventable conditions \cite{Arozullah_2003}. 
Those with poor health literacy skills may also be more vulnerable to false information about the pandemic and have trouble evaluating the quality of information sources compared to those with higher levels of health literacy \cite{paakkari_covid-19_2020}.

The pandemic has highlighted racial inequalities, particularly in the United States. 
Americans of Hispanic and Latinx decent were found to represent 28.7\% of the cases nationwide while accounting for 18.45\% of the total US population. 
Hispanic/Latinx and Black non-Hispanic people also represent more deaths than their share of the population would suggest (Hispanic/Latinx: 18.7\% deaths vs. 18.45\% of population, Black: 13.7\% vs. 12.54\%) \cite{CDC_tracker}. 
Additionally, there is evidence that Black people in the US are more likely to be infected, hospitalised, and die of COVID-19 compared to their White counterparts \cite{mackey_racial_2021, price-haywood_hospitalization_2020}.

Essential workers were especially at risk of contracting COVID-19 in the first two waves, given that they cannot self isolate like other workers. 
Essential workers comprise about 40\% of the US work-force, and include industries such as healthcare, social services, food services, cleaning, transit, essential manufacturing, and agriculture. 
Studies suggest that essential workers are at higher risk of being infected with COVID-19 compared to the general population due to increased proximity to others \cite{milligan_impact_2021, cox-ganser_occupations_2021}. 
Moreover, 25\% of US essential workers were estimated to be low income, and 11\% were estimated to lack health insurance \cite{mccormack_economic_2020}, putting them at even greater risk due to the aforementioned risks associated with poverty and being uninsured.
A study conducted in the United Kingdom found that essential workers were at higher risk of severe disease and mortality due to COVID-19 compared to non-essential workers \cite{mutambudzi_occupation_2020}. 

In this paper we focus on the impact of COVID-19 on Californian counties and their demographic features. 
Studies which are similar in design tend to focus on all US counties, such as a study linking overcrowding and COVID-19 incidence and mortality rates \cite{kamis_overcrowding_2021}, rather than a single state or have linked environmental and demographic factors to increased risk of transmission and mortality at the individual level, but not as they pertain to the population level. 
The goal of this study is to measure the correlation of the demographic features in terms of how they effect population at a large scale rather than the individual. 
Therefore, features that represent a marginal set of the population may not have a significant correlation to the outcomes compared to other features which describe or effect larger sections of the population.

General measures of a community's vulnerability to disease outbreaks such as COVID-19 do exist. 
The California Healthy Places Index (HPI) describes the health of Californian communities based on social determinants of health \cite{HPI}. 
The Social Vulnerability Index (SVI) by the CDC and Agency of Toxic Substances and Disease Registry (ATSDR) is a more general measure that determines a community's vulnerability to emergencies, such as natural disasters and disease outbreaks, based on its socioeconomic status, housing and transport conditions, racial and ethnic background, and household composition \cite{ATSDR}. 
This study aims to compare some of the features also included in these vulnerability measures together to see which are the most impactful for COVID-19 specifically, not in terms of general health or disaster preparedness.
More specifically, this study seeks to find which of these features are most correlated with COVID-19 cases and deaths in California specifically. 
While social inequities apply generally across the entire United States, this study may be used as a point of comparison for similar studies done either nationwide or for other states.

\section{Material and Methods}

COVID-19 data on daily cases and deaths by county were sourced from the COVID-19 Time Series Metrics dataset from the Vaccine Progress Dashboard data on the California Open Data Portal \cite{Cal_OpenPortal} on July 27 2021 available from which cases and deaths per 100,000 were calculated.

Most demographic data was sourced from the United States American Community Survey (ACS) Data  \cite{ACS} and specifically from the 2019 1-Year Estimates. 
While 2020 data were available, 2019 demographic data was used for this analysis since it is more representative of what communities were like approaching the pandemic and 2020 values were not available for some datasets. 
Age group data (in which most age groups are split into groups spanning 5 years except for the group of 85+ years) and race/ethnicity data was obtained from the US Census website \cite{US_Census}. 
All other demographic characteristics, namely population employed in various industries, education attainment, disabled population (civilian, non-institutionalized), health insurance status (private, public, or uninsured), language spoken at home, English ability, median income, and average family size, were downloaded through the use of the US ACS API for 2019 1-Year data instead \cite{ACS}. 
The reference populations for each group of variables are listed in Table \ref{tab:refPop}.

Demographic data were merged by county, then merged with COVID-19 data (restricted to the most recent date at the time of analysis, 5 July 2021). 
Variables of cumulative cases per 100,000 and cumulative deaths per 100,000 were created using the population and cumulative case count and cumulative deaths count features within the newly merged data frames, and will be the variable of interest for this analysis.
Counties missing census data due to low population figures ($<65,000$ people) were removed for two key reasons: 1) they were missing a substantial amount of demographic data and 2) low population counties are more likely to be subject to high variability in COVID-19 spread compared to more populous counties. 
The following counties were removed: Del Norte, Siskiyou, Modoc, Trinity, Lassen, Plumas, Glenn, Sierra, Colusa, Alpine, Amador, Calaveras, Tuolumne, Mono, Mariposa, San Benito, and Inyo, leaving 41 counties remaining in the analysis.

Race and ethnicity data contains the groupings: Asian, American Indian and Alaska Native, Black, White, Native Hawaiian and Other Pacific Islander, Hispanic, and Two or More Races. 
Here, `Hispanic' refers to people of Spanish or Latin-American origin aside from Brazilian. 
This group overlaps with all racial categories, as Hispanic people may be of any race. 
All other categories besides Two or More Races refer to all people who identify as that race, including those of two or more races.

Industry values are determined based on the job full-time, year-round employed civilians have worked at the most in the past week. For unemployed civilians who have worked in the past 5 years, this refers to the industry of the job previously held.
Employment status categories sample the total population of those 16 years and older. 
The list of features here includes enlistment in the armed forces as a percent of this population in addition to civilian employment, which may lower employment rates in areas with high numbers of military personnel \cite{ACS}.

Some data groupings were merged. 
Age data was regrouped from groups of 5 years to the following groups: 0-19, 20-39, 40-49, 50-59, 60-69, 70-79, 80+. 
Race/ethnicity and industry data were originally subdivided by gender, but were summed together to create a total value. 
In the case of age and race/ethnicity, percentages of total population were attained for both the age and race/ethnicity variables by dividing each variable by the total population for each county. 
This was not needed for the industry data since the percentage values could be summed from the percentages sourced from the API.

Housing overcrowding data by county was also obtained from the California Open Data Portal \cite{CalOpenData_Overcrowding}. In this dataset, overcrowding is defined as a housing density greater than 1.0 persons per room and severe overcrowding is defined as housing density greater than 1.5 persons per room. Data from 2019 for county-wide populations were used. 
Data constrained to 2019 was extracted and only county, overcrowding, and severe overcrowding estimates (as percentage of households) were retained.

Preventable hospitalisation data was downloaded from the California Open Data Portal \cite{Cal_OpenPortal}. 
This data used from this data set were the county-wide 2019 observed rates (per 100,00 of population) of preventable hospitalisations related to Diabetes (Short-term complications, long-term complications, uncontrolled, and lower-extremity amputations among patients with diabetes), Asthma in Younger Adults (Ages 18-36), COPD or Asthma in Older Adults (Ages 40+), Hypertension, Community-Acquired Pneumonia, and Urinary Tract Infections \cite{CalOpenData_PrevHospitalisations}.

Voter registration data was sourced from Statement of Vote: General Election November 3, 2020 and voter registration numbers were obtained from the ``Report of Registration as of October 19, 2020" \cite{Cal_GeneralElection}.
Republican and Democrat party estimates were acquired for each county by dividing total people registered in each party for that county by the total people registered to vote in that county.

Features that fell under the following categories were removed from the dataset: a) features missing data, b) racial features not defined as `percent of people of (a race) alone or combination with another race', c) daily data that changes over time (such as temperature data, day of the week, and mobility data), d) all other COVID-19 data, e) raw number features already represented by a percent of the population, and f) all other redundant features (such as working population over 16 years).

Pearson correlation coefficients ($r$-values) as well as $r^2$ values were calculated between all population features, cumulative cases per 100,000, and cumulative deaths per 100,000 as of 5 July, 2021. 
Feature pairs and their corresponding correlation values were filtered to find the highest and lowest correlation values for each feature, by sorting by $r^2$ in the appropriate direction.
For the purposes of describing correlations, a pair of features will be considered `strongly' correlated if the absolute value of their correlation coefficient is greater than 0.5 ($\vert r \vert \ge 0.5$), moderately correlated if the absolute value is between 0.5 and 0.2 ($0.5 \ge \vert r \vert \ge 0.2$), and weakly correlated if the absolute value is between 0.2 and 0.1 ($0.2 \ge \vert r \vert \ge 0.1$).
Any correlation whose absolute value is less than 0.1 ($\vert r \vert \le 0.1$) is considered to be almost non-existent. 
Note that while it is fact that the closer the absolute value of $r$ is to 1, the closer the relationship between the features resembles a linear relationship.

We created 4 linear models to further quantify the strength of the relationship between the most highly correlated features and our target features, cumulative cases and cumulative deaths per 100,000, especially when multiple input features are used within the same model. This approach specifically was used due to the ease of implementation compared to other models.

Using these features, two linear regression models were created. 
The target variable and feature sets were assigned to each model as shown in Table \ref{tab:models}.
\begin{itemize}
\item Model 1 contains the top 5 most significant features to cumulative cases per 100,000. 
\item Model 2 uses the same target variable (cumulative cases per 100,000), but attempts to minimise co-linearity by selecting from a pool of the top 20 features by $F$ regression test which are un-likely to be co-linear with each other. 
\end{itemize}
To meet this requirement, all input features in Model 2 must have a Pearson correlation of less than 0.5 with all other input features. 
These features were selected manually according to this rule. 
Models 3 and 4 follow the same procedures, but for the target variable of cumulative deaths per 100,000 instead.

We verified the linearity of each model by testing:
\begin{enumerate}
\item \textbf{that the relationships between the target variable and each input variable are linear in nature} by performing the Linear Rainbow test for linearity, which tests if the `middle` 50\% of residuals fit significantly worse than the whole distribution. ($H_0$ : The middle 50\% of residuals do not differ significantly in fit compared to the entire set of residuals; $H_1$ : The middle 50\% of residuals overall fit is significantly worse than for the entire set of residuals.) 
This test assumes homoscedasticity from the residuals, and may reject an otherwise linear model if assumption 3) is violated \cite{obrien_caution_2007}.

\item \textbf{that residuals are independent of each other} was assessed by calculating the Variable Inflation Factors (VIF) values for the input variables of each model. 
A rule of thumb is that if any of these values exceeds 5, then there is likely a linear relationship between at least two of the input variables, which would violate this assumption \cite{obrien_caution_2007}.

\item \textbf{that residuals are homoscedastic, or that they have a constant variance for all values of the target variable} by performing the Breusch-Pagan test for heteroscedasticity on the residuals of a linear regression model. ($H_0$ : The variance of errors from the model are homoscedastic, $H_1$: The variance of errors from the model are *not* homoscedastic, but rather heteroscedastic.) \cite{breusch_simple_1979}. 
The Koenker variant of this test as it does not assume the residuals follow a normal distribution \cite{koenker_note_1981}.

\item \textbf{that the residuals follow a normal distribution \cite{poole_assumptions_1971}} by performing the Jarque-Bera test for normality of residuals. The test detects if the residuals of the linear model have the skewness and kurtosis that match a normal distribution, with the null hypothesis stating that the distribution is normal. ($H_0$ : The distribution of the residuals follows a normal distribution, $H_1$ : The distribution of the residuals does *not* follow a normal distribution) \cite{jarque_efficient_1980}.
\end{enumerate}

The significance level for each of these tests was set at 0.95 (or $\alpha$ = 0.05).

All models were evaluated for overall significance using the $F$-test to measure if the models perform statistically significantly better than a model with no input variables ($H_0 : \beta_1 = \beta_2 = \ldots = \beta_i$ where $\beta$ signifies a coefficient in the linear model and $i$ number of input variables vs. $H_1 : \beta_j \ne 0$ for any $j$ integer in $[0, i]$) and the $t$-test for the significance of each input variable to determine the significance of a given input variable in the model ($H_0 : \beta = 0, H_1 : \beta \ne 0$ for $\beta$ coefficient of a given input variable in the linear model).

Input features from all models were evaluated for other demographic features they were strongly correlated with (defined as $\vert r \vert \ge 0.50$ between the feature pair). These relationships will be defined as co-correlations or co-correlated features. The procedure for this was to create a function that iterates through the list of each models features and counts the number of times each feature outside of that set is strongly correlated with each item in that list.

\section{Results}

After filtering counties for which demographic data was missing, 41 counties remained. For the remaining counties, cumulative cases per 100,000 and cumulative deaths per 100,000 had means of 8068 and 119 (rounding to the nearest whole number), with standard deviations of 2788 and 66, and IQR ranges of 5947 to 9957 and 73 to 156 respectively. 

\begin{figure}[ht!]
\begin{center}
\includegraphics[width=17cm]{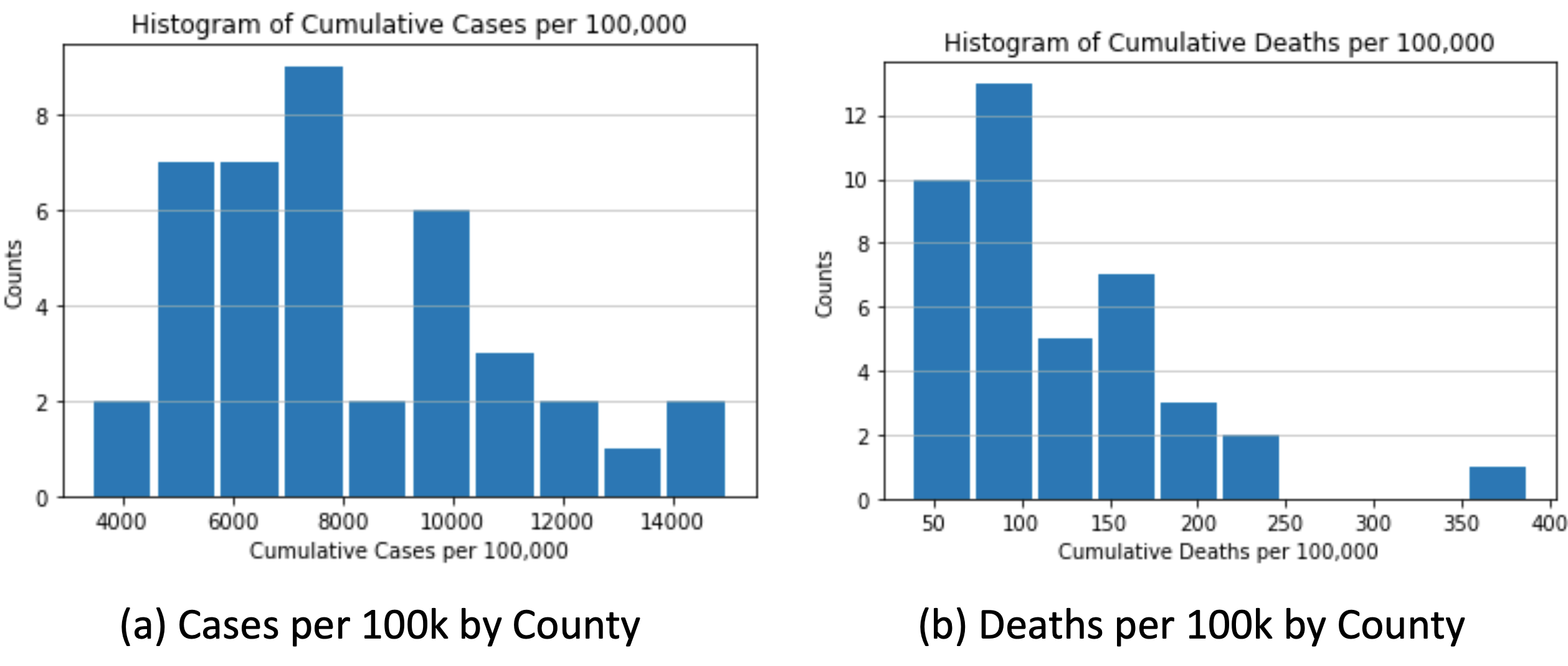}
\end{center}
\caption{Histograms of Cases (left) and Deaths (right) per 100,000 by California Counties. The outlier in the figure on the right is Imperial County.}
\label{fig:cumCasesAndDeaths}
\end{figure}

\begin{figure}[ht!]
\begin{center}
\includegraphics[width=17cm]{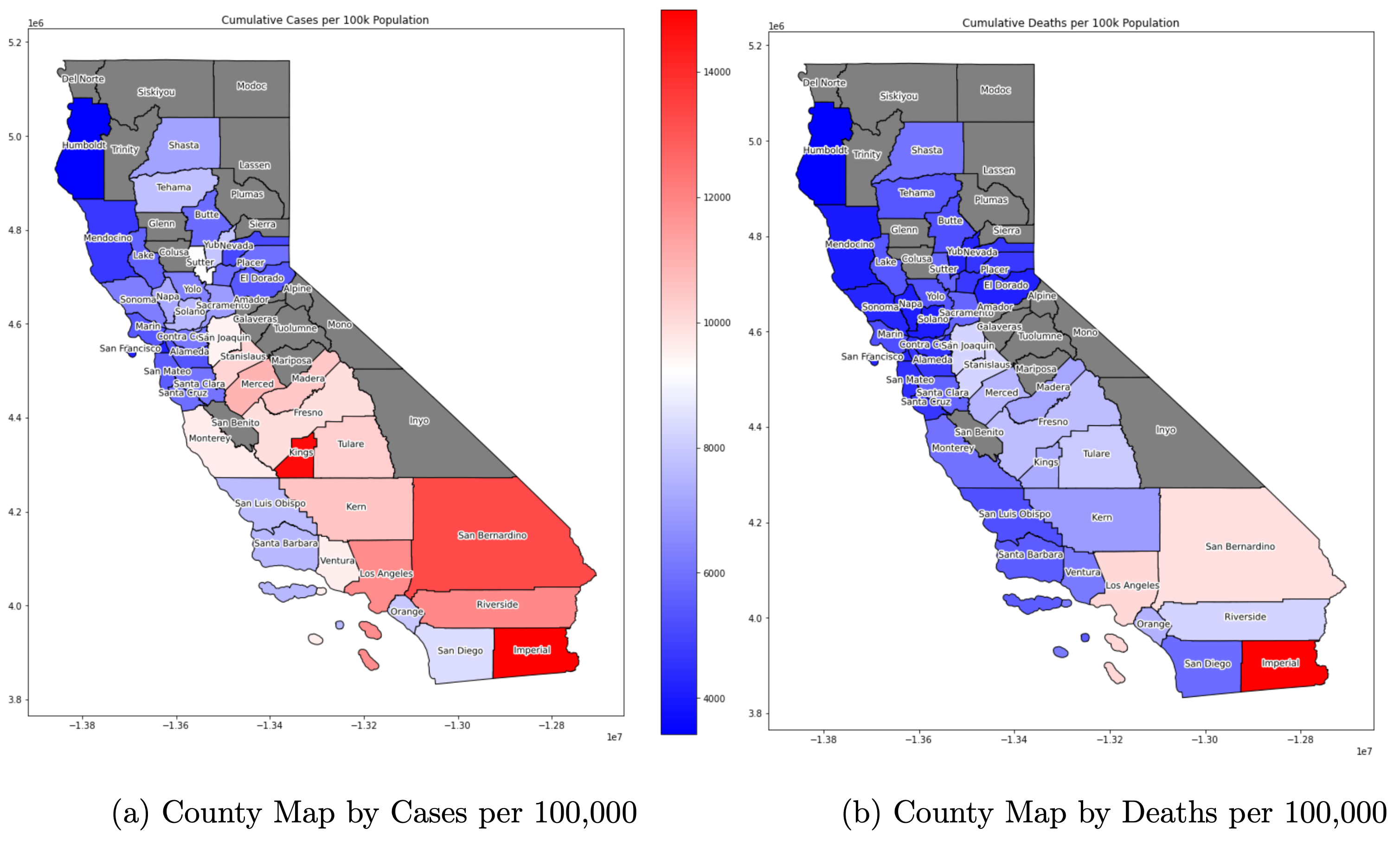}
\end{center}
\caption{County Map by Cases and Deaths per 100,000 of the population.}
\label{fig:countyCasesAndDeaths}
\end{figure}

The histograms of cumulative cases (Figure \ref{fig:cumCasesAndDeaths}a) and deaths (Figure \ref{fig:cumCasesAndDeaths}b) per 100,000 are skewed towards larger numbers of cases and deaths with one clear outlier (Imperial county) in the cumulative deaths plot. According to the county maps for cases and deaths (represented in Figures \ref{fig:countyCasesAndDeaths}a and \ref{fig:countyCasesAndDeaths}b respectively), cases and deaths in southern counties was higher compared to northern counties. Furthermore, coastal counties tend to have better outcomes than inland counties.

The counties with the top 5 cumulative cases per 100,000 values were (in descending order) Imperial, Kings, San Bernardino, Riverside, and Los Angeles counties, and the bottom 5 (in ascending order) were Humboldt, San Francisco, Mendocino, Nevada, and Alameda counties. For cumulative deaths per 100,000, the counties with the top 5 highest values were Imperial, Los Angeles, San Bernardino, Inyo, and San Joaquin counties, with the bottom 5 being Mariposa, Mono, Trinity, Del Norte, and Plumas counties.

\subsection{Correlations with Cases and Deaths per 100,000}

\paragraph*{Race, Ethnicity and Sex}
The percentage of female population correlates moderately with fewer cases per 100K (-0.480) and fewer deaths (-0.242). For males, the opposite is true (0.480, 0.242).

Most racial categories had moderate to weak correlations with cases and deaths per 100K with exception of percentages of individuals of two or more races and Hispanic ethnicity. Hispanic ethnicity strongly positively correlates with cases and deaths (0.875, 0.795). Two or more races correlates negatively and moderately for cases and deaths (-0.494, -0.557). Asian (-0.267, -0.167), Native Hawaiian or Pacific Islander (-0.186, -0.228), and American Indian or Alaska Native (-0.092, -0.129) populations had moderate and weak correlations with cases and deaths at best. Black (0.257, 0.135) populations had a moderate correlation with cases and weak correlation with deaths while White (0.129, 0.078) populations had weak positive correlations with cases but essentially no correlation with deaths.

\paragraph*{Age}
Younger populations are linked with increases in cumulative cases and deaths while older populations are linked with a decrease in these outcomes. The 0-19 group is strongly correlated with both of these variables (0.752, 0.604) and a substantially weaker correlation exists for the 20-39 group (0.270, 0.210). On the other hand, age groups of 50-59 or greater are strongly associated with fewer cases and deaths per 100,000 (-0.492, -0.390).

\paragraph*{Education Level}
In general, higher levels of education correlate with fewer cumulative cases and deaths while the opposite is true for lower levels of education. Categories of ``less than 9th grade", ``9th to 12th grade, no diploma", ``High school graduate (or equivalent)", and ``Some College, no degree" all correlate positively with cases (0.683, 0.747, 0.443, 0.101 respectively). The first three of these features correlate positively with cumulative deaths (0.528, 0.632, 0.346 respectively) with ``Some College" having essentially no correlation with this variable (0.015). ``Associate's degree", ``Bachelor's degree", and ``Graduate or professional degree" all correlate to fewer cases with the relationship being strongest for the last two education levels (-0.256, -0.591, -0.580). Similar relationships can be found between these features and cumulative deaths (-0.326, -0.418, -0.444).

\paragraph*{Health Insurance Status}
Greater proportions of people possessing health insurance correlates negatively with COVID-19 cases and deaths (-0.647, -0.517) as does the proportion of people possessing private health insurance (-0.577, -0.570). However, the correlation between cases and deaths becomes moderately positive for the proportion of those with public health insurance (0.326, 0.341).

\paragraph*{Disability Status}
The proportion of non-institutionalized civilians with a disability correlates positively with COVID-19 cases and deaths (0.513, 0.399).

\paragraph*{Preventable Illness Incidence Rates}
Most preventable illnesses were found to have a slight positive correlation with COVID-19 cases. Long-term diabetes complications were found to have the highest correlation with COVID-19 incidence, followed by instances of lower-extremity amputations (0.691, 0.490) and a similarly high correlation with deaths per 100,000 (0.578, 0.267). Short-term diabetes complications, uncontrolled diabetes, community-acquired pneumonia, hypertension, and urinary tract infections had less strong correlations to cases (0.223, 0.299, 0.209, 0.257, 0.346) and similar less strong correlations with deaths (0.072, 0.190, 0.236, 0.252, 0.351). Heart failure, asthma in younger adults (ages 18-39), and COPD or Asthma in Older Adults (ages 40+) had very weak to no correlation with cases (0.188, -0.056, 0.021) and deaths (0.143, 0.072, 0.013).

\paragraph*{Industry Employment}
While most sectors had weak or moderate correlations, a few stand out as having stronger correlations to COVID-19 cases. ``Agriculture, forestry, fishing and hunting, and mining" (0.536, 0.306), ``Transportation and warehousing, and utilities" (0.631, 0.509), and ``Whole-sale trade" (0.541, 0.481) had the strongest positive correlations with cumulative cases and deaths whereas ``Finance and insurance, and real estate and rental and leasing" (-0.500, -0.344) and ``Professional, scientific, and management, and administrative and waste management" (-0.488, -0.348) had the strongest negative correlations. Industries with moderate-to-weak correlations include ``Public administration" (0.314, 0.187), which had positive correlations with cases and deaths and ``Arts, entertainment, and recreation, and accommodation and food services" (-0.308, -0.175), ``Educational services, and health care and social assistance" (-0.202, -0.123), ``Information" (-0.352, -0.171), and ``Other services, except public administration" (-0.206, -0.127), which were negatively correlated to cases and deaths. ``Construction" (-0.013, 0.045), ``Manufacturing" (-0.012, -0.066), and ``Retail trade" (0.031, 0.060) had virtually no correlation with cases or deaths.

\paragraph*{Commute Methods and Duration}
Commuting by car via carpooling (0.354, 0.209) and alone (0.486, 0.339), was found to be positively correlated with cumulative cases and deaths. Working from home (-0.601, -0.434) was found to have the strongest negative correlation with cases and deaths out of all features describing commute methods. However, walking (-0.385, -0.242), public transportation (-0.358, -0.227), and ``other means" (-0.134, -0.120) were all found to have moderate-to-weak negative correlations with cases and deaths. This is a surprising result that will be explored in a later section. Mean travel time to work had no correlation with either cases or deaths (0.019, 0.034).

\paragraph*{Language Proficiency}
Language proficiency (or lack thereof) and whether or not English was the only language spoken at home had almost no correlation with COVID-19 incidence. Proportion of those who only speak English ``less than \textit{very well}" (0.016, 0.053) as well as those who speak only English at home (0.016, 0.053) had both had no correlation to cases or deaths.

\paragraph*{Party Affiliation}
Party affiliation had no significant correlation with cumulative deaths and only a small correlation to cases per 100K with the correlation of registered Democrats (-0.201, -0.090) and Republicans (0.264, 0.097).

\paragraph*{Overcrowding and Family Size}
Average family size (0.816, 0.777) and overcrowding (defined as $>1.0$ people per room) (0.681, 0.593) correlate strongly with both cases and deaths. There is a moderate correlation between severe overcrowding ($>2.0$ people per room) (0.361, 0.347) and cases and deaths, respectively.

\paragraph*{Residence Continuity}
The proportion of people who lived in the same residence a year ago (0.353, 0.387) and the proportion of those who lived in a different residence in the same county a year ago (-0.164, -0.057) weakly correlate to cumulative cases positively and negatively respectively. However, the correlation with deaths only exists for the former with no correlation with the latter variable.

\subsection{Linear Models}

We take the 5 most highly correlated features and create a linear model to statistically model infection and death rates within California.

Model 1 is composed of the following features: 
\begin{itemize}
\item Percent of population identifying as Hispanic ethnicity (r = 0.875), 
\item Average family size (0.816),
\item Age 0-19 (as percent of population) (0.752), 
\item Education level of 9th to 12th grade, no diploma (Percent $\ge5$ years) (0.747), and
\item Incidences of Long-term Diabetes Complications (per 100K of population) (0.690). 
\end{itemize}
The overall model was found to be significant by F-test of overall significance $(F = 29.12, p >0.001)$. Out of the selected input variables, only Hispanic ethnicity was found to be significant at significance level of $\alpha = 0.05\ (t = 2.473,\ p = 0.018)$.

While the  input variables in this model are not mutually co-linear, calculating the Variance Inflation Factor (VIF) of each variable show that Hispanic ethnicity and average family size have VIF values equal to 7.595 and 5.269 (VIF values greater than 5 suggests high collinearity with other input variables \cite{obrien_caution_2007}), we can assume that this model does a poor job of establishing correlations between each separate input variable to the target variable of cumulative cases per 100K.

However, Model 1 did exceed the level of significance required to accept the corresponding alternative hypotheses for Jarque-Bera test for normality, the Breusch-Pagan test for heteroscedacity, or the Linear Rainbow test for linearity at our chosen significance level $(p = 0.470,\ 0.114,\ 0.474 \text{ respectively at } \alpha = 0.05)$.

A second model, Model 2, is composed of the following features: 
\begin{itemize}
\item Overcrowding (percent of households) (r = 0.681), 
\item Proportion of employed population working in the wholesale trade industry (0.541), 
\item Proportion of employed population working in the industries of transportation, 
\item Warehousing, and utilities (0.631), and
\item Proportion of population over the age of 25 with a graduate or professional degree (-0.580). 
\end{itemize}
The overall model was found to be significant by F-test of overall significance $(F = 35.32, p < 0.001)$. The variables of overcrowding and transportation, warehousing, and utilities, and graduate education were all found to be significant at significance level of $\alpha = 0.05 (t = 5.288, p < 0.001, t = 4.384, p < 0.001, \text{ and } t = -2.452, p < 0.019 \text{ respectively})$.

There was not significant evidence to suggest that Model 2 violated any assumptions of the linear model. It did not fail the Jarque-Bera test for normality, the Breusch-Pagan test for heteroscedacity, or the Linear Rainbow test for linearity at our chosen significance level. $(p = 0.352, 0.276, 0.547 \text{ respectively})$. The VIF values for each input feature for Model 2 never exceed 5, but this assumption is violated, therefore it is reasonable to assume that multicollinearity is minimal in our model, which is to be expected given the way our input features were selected.

Model 3 is composed of the following features:
\begin{itemize}
\item Overcrowding (percent of households) (r = 0.593), 
\item Hispanic ethnicity (0.795), 
\item Average family size (0.777), 
\item Age 0-19 (as percent of population) (0.604), and
\item Education level of 9th to 12th grade, no diploma (Percent >= 25 years) (0.632). 
\end{itemize}
The model was found to be significant overall by F-test of overall significance $(F = 17.05, p < 0.001)$. Only Hispanic ethnicity and average family size were found to be significant features according to $t$-test $(t = 2.713, p = 0.010 \text{ and}, t = 2.086, p = 0.044 \text{ for each feature respectively})$.

Model 3 mostly appears to follow the assumptions of the linear model except for the assumption of independence between input variables. It did not fail the Jarque-Bera test for normality, the Breusch-Pagan test for heteroscedacity, nor the Linear Rainbow test for linearity at our chosen significance level $(p = 0.600, 0.216, 0.269 \text{ respectively})$. However, the VIF values for ``Ages 0-19" and ``Average family size" are 9.568 and 4.755 respectively, suggesting these variables are highly collinear with each other.

Model 4 is composed of the following features: 
\begin{itemize}
\item Overcrowding (percent of households) (r = 0.593), 
\item Proportion of employed population working in the wholesale trade industry (0.481), 
\item Proportion of employed population working in the industries of transportation, 
\item Warehousing, and utilities (0.509), and 
\item Proportion of population over the age of 25 with a graduate or professional degree (?0.444). 
\end{itemize}
The overall model was found to be significant by F-test of overall significance $(F = 11.53, p > 0.001)$. By $t$-test, only overcrowding and transportation and warehousing, and utilities were found to be significant features $(t = 3.244, p = 0.003 \text{ and } t = 2.511, p = 0.017 \text{ respectively})$.

However, while the model was found to be significant, it fails two of the four assumptions of a linear regression model. It fails the Jarque-Bera test for normality of residuals $(p < 0.001)$ as well as the Linear Rainbow test $(p = 0.011)$. The Breusch-Pagan test failed to reject the assumption that the data were homoscedastic $(p = 0.471)$. None of the VIF values for these variables exceeded 2, indicating low collinearity in this model.

Correlations between the Model 1's features overlap substantially. In addition to features being correlated, the following variables were co-correlations to all input features: overcrowding, Ages 60-69, Ages 80+, health insurance status (none or any), private health insurance, percent of employed working from home, and education level less than 9th grade. Co-correlations with at least 4 of the input features include the age groups of 50-59 and 70-79, the ``Agriculture, forestry, fishing and hunting, and mining" industry group, Bachelor's degrees, the ``Finance and insurance, and real estate and rental and leasing" industry group, and unemployment rates.

Co-correlations between Model 2 and 4's features and the main feature set have less overlap with each other than Model 1. Features co-correlated with 3 input features are 9th to 12th grade education, no diploma, long-term diabetes complications, and Ages 0-19. Features strongly correlated with 2 input features are proportion of those who work from home as their commute, health insurance status (any and none), the ``Agriculture, forestry, fishing and hunting, and mining" industry group, the high school graduate (or equivalent) education, proportion of disabled population, average family size, and proportion of population identifying as Hispanic.

For Model 3, the features have many co-correlations in common with Model 1 given that both have a similar feature set. The features which co-correlate with all input features for Model 3 were long-term diabetes complications, working from home, health insurance status, age groups of 60-69 and 80+, and less than 9th grade education level. Co-correlated features with at least 4 of the input features were private health insurance, ages 50-59, ages 70-79, and ``Agriculture, forestry, fishing and hunting, and mining" industry group. %Other correlated features can be found in Table \ref{TODO}.

This evidence points to a strong correlation between cumulative cases and deaths per 100,000 and the conditions of poverty and low socioeconomic status, high levels of employment in essential work, and racial inequalities.

\section{Discussion}

This study observed a variety of demographic features simultaneously and showed that correlations suggest a strong link between COVID-19 spread and housing size and density, low education attainment, low access to health services, employment in specific industries, and minority status, consistent with existing literature on health outcomes as it pertains to social status \cite{wang_airborne_2021,greenhalgh_ten_2021, marmot_health_1991}. COVID-19 spreads best indoors and where the infected are in close proximity to others \cite{wang_airborne_2021,greenhalgh_ten_2021}. Those who live in especially crowded conditions may lack the ability to self-isolate from family and housemates at all. Low levels of education attainment correlates with fewer, more lower paying work opportunities. Lacking health insurance is both an indicator of a lack of means to afford it and can lead to situations where care is accessed at later stages of illness, such as for diseases like diabetes. It is also well known that higher income and status leads to better health outcomes overtime, due to greater ability to meet one's own needs and decrease or eliminate poverty-related stressors \cite{marmot_health_1991}.

Many of the industries that positively correlated with COVID-19 spread were related to sectors considered essential, such as agriculture, wholesale trade, transportation and warehousing, utilities, and similar industries. On the other hand, industries that had strong negative correlations with COVID-19 spread included finance, insurance, real estate, information, professional, scientific, and management jobs. These industries are not considered essential and many of these positions would have been most likely been reconfigured into remote positions. This suggests essential workers in essential industries are more likely to be at risk of contracting COVID-19 since they would not have been able to do their work remotely. This simple conclusion from a rather simple data analysis hides an important feature of human behaviour; those at perceived greater risk may take more precautions. For example retail trade had negligible correlations with cases and deaths, perhaps due to the mandated mask wearing and hand washing, while agriculture, hunting, forestry being outdoor professions may have a lower perceived risk and, as a result, lower preventative measures resulting in higher correlation with cases and deaths.

\begin{figure}[ht!]
\begin{center}
\includegraphics[width=15cm]{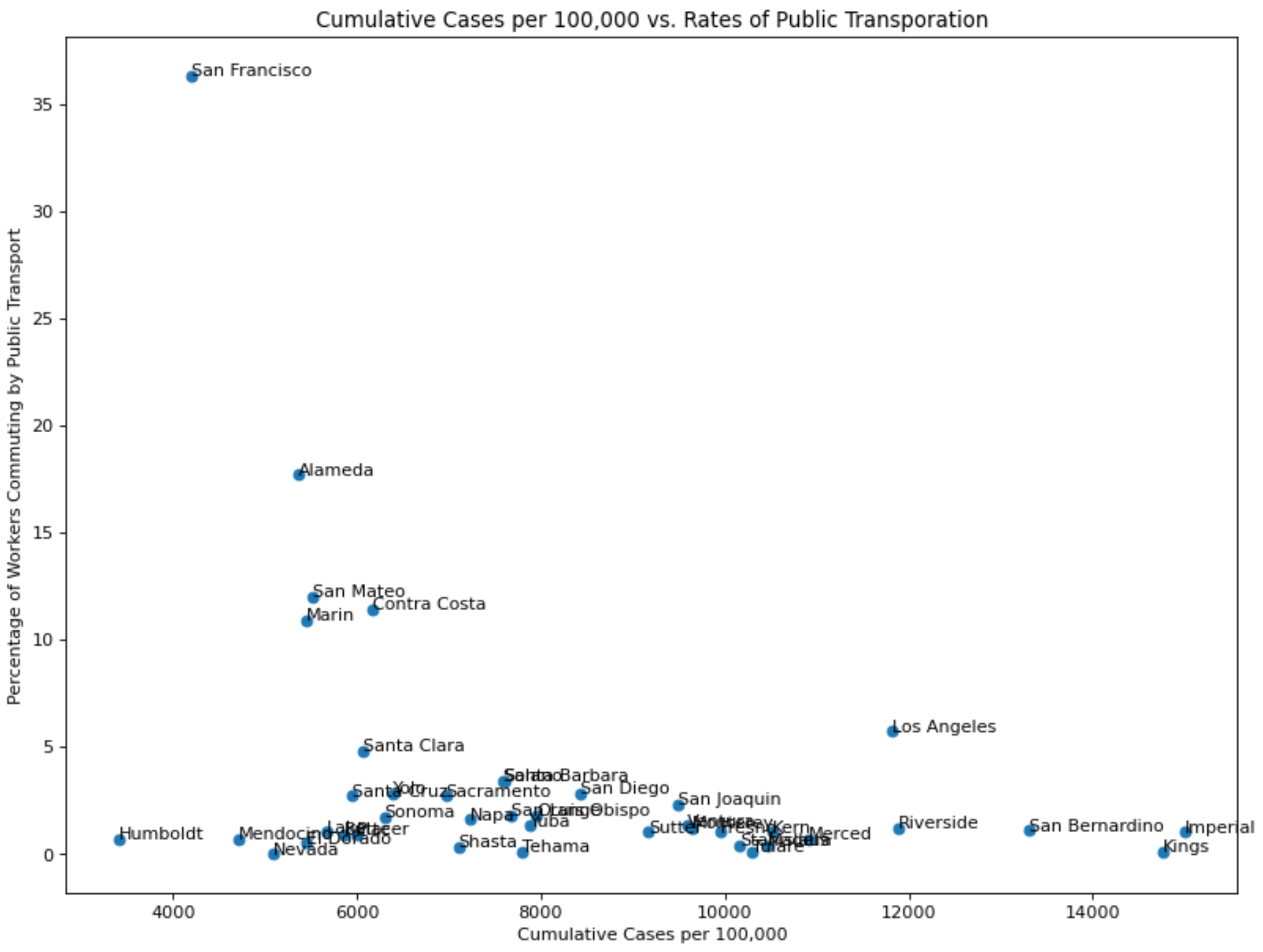}
\end{center}
\caption{Graph of Cumulative Cases per 100,000 vs. Percentage of Workers Commuting by Public Transportation}
\label{fig:cumDeathsCommuting}
\end{figure}

The use of the Pearson Correlation as the main metric for this study implicitly assumes any correlation will follow a linear relationship with few (if any) outliers. This is was not always the case for some variables. For example, the $r$ value for commuting by public transportation suggests there is a moderate negative correlation (-0.358). However, not only does this not line up with research  \cite{wang_airborne_2021,heinzerling_transmission_2020}, but it is apparent in figure \ref{fig:cumDeathsCommuting} that the correlation is being driven by a few outliers in the Bay Area (especially San Francisco). When eliminating these outlier counties, there appears to be almost no correlation between the use of public transportation and COVID-19 outcomes which seems counter-intuitive at first but personal precautions and lower levels of public transport use may explain this result.

Secondly, this study is limited in its ability to isolate the effects of every variable from each other. The correlation between higher proportions of people age 0-19 and higher incidences of cases and deaths seems like a contradiction considering that older individuals are typically at greater risk of severe illness and mortality. However, this age group correlates positive with larger average family sizes and the prevalence of over-crowded households. Therefore, it is likely that the more people ages 0-19 there are in a population (relative to other age groups), the larger households are on average and more crowded homes there are on average. This means that children do not cause worse COVID-19 outcomes per say, but that the proportion of children in a population correlates with other features that are more inherently harmful from a public health standpoint. This is an important point as the features we determine from correlations with cases and deaths may actually be proxies for other features that do  not show up so prominently in analysis. 

The correlation between certain groups within the population and COVID-19 outcomes become less reliable as said groups make up smaller portions of the population. For example, the interquartile range (IQR) for the percentage of White and Hispanic populations in a given county are 76.9\% to 90.0\% and 22.9\% to 48.6\% respectively. Meanwhile, Black, Native American/Alaska Native, and Native Hawaiian/Pacific Islander populations have much narrower IQRs in addition to making up a smaller share of county populations overall. This means that conclusions based on correlations from these population percentages are limited to this narrow range of percentages. Furthermore, these correlations may be more descriptive of counties where people of these groups tend to make up more of the relative population rather than describe the actual correlation of racial populations to COVID-19 cases, especially if said populations represent a fraction of a relatively small percent of most counties.

Furthermore, demographic data for each county is sourced from 2019, meaning that these data do not capture the effects the pandemic had on the population after the pandemic took hold in the United States. Some demographic data, such as unemployment levels, would have changed substantially throughout the pandemic. However, this was a deliberate choice for most of the data since the goal of the study was to see how these features would effect COVID-19 outcomes dependent on how counties appeared before the onset of the pandemic. This is also simpler than attempting to account for changes in features, such as unemployment.

The features determined in this study suggests that vaccination strategies must prioritize the most disadvantaged and essential workers if the disease is to be brought under control. Medi-Cal (Californian public insurance for low-income, non-elderly individuals) recipients were to receive monetary incentives for receiving their vaccinations to boost vaccination rates among low-income populations though this program does not account for the uninsured and a substantial number of low income, uninsured individuals did not currently qualify for these incentives. Vaccination strategies should also include efforts to educate everyone on the safety of the vaccine in a way that is inclusive to these groups. In a  nationwide poll by the Kaiser Family Foundation, respondents that reported not getting the COVID-19 vaccine reported side effects and the newness of the vaccine as the main reason for their decision respectively but many of these respondents were not resistant to getting a vaccine in the future. This suggests a need to communicate to the public that the vaccines are safe and that the risk of side effects outweigh the risk of the virus. Considering that many people in poverty may not be able to miss a day of work or access alternative childcare, efforts must also be sensitive to these concerns.

Measures to assist those in overcrowded housing to self-isolate may also be beneficial. An Australian study suggested that the strategy of isolating COVID-19 patients in hotel rooms rather than their homes becomes more cost effective to the state as the patients household size increases and/or if the household has older individuals \cite{effectiveness_2021}. While cost effectiveness analysis does not apply in the United States, which lacks a universal healthcare system, a similar strategy may still be effective in reducing the burden on hospitals, protecting larger households, and reducing the number of people who will bear the burden of medical costs due to their COVID-19 related hospital stays.

Since the main limitation of this study was the inability to directly isolate the effects of specific features on how COVID-19 impacted each county, further research is needed to accomplish this very goal. An example study may involve an observational study aiming to determine if the size of a household or family increases risk of COVID-19 illness and death. This study would ideally control for socioeconomic status and overcrowding within the household. A similar study could be done to determine the effect of COVID-19 on racial and ethnic minorities to determine if simply being a specific minority puts one at increased risk of COVID-19 illness and mortality even independent of racial and ethnic disparities in wealth.

Further investigation of Imperial County COVID-19 mortality rates may be beneficial in determining why this county is such an outlier in this regard, even when accounting for demographic features. A hypothesized cause for this is that a substantial number of people (Mexican and American citizens alike) that commute between the county and nearby Mexicali in Mexico [60]. This would increase the risk of transmission in times when Mexicali was experiencing their own COVID-19 spike. Moreover, it may be that a substantial number of American citizens living in Mexicali (14 such cases were cited during a hospital surge at El Centro Hospital in May 2020 [61]) sought medical care in Imperial County, placing hospitals under even more pressure relative to population. Further research would be needed to clarify which factors drove COVID-19 mortality and to what extent.

\begin{table}[htp]
\caption{Reference Populations for Data used in this study. *Estimate includes the industry category of those who have been employed within the last 5 years.}
\begin{center}
\begin{tabular}{|l|l|}
\hline
Feature & Sampled Population \\
\hline
Race/Ethnicity  & Total Population \\
Sex  & Total Population \\
Age Categories & Total Population \\
Registered Party Affiliation & Registered Voters \\
Education Level & Population $\ge$ 25 years \\
Employment Status  & Population $\ge$ 16 years \\
Health Insurance Status  & Civilian Non-institutionalized Population \\
Industry & Full-time, year-round civilian employed population $\ge$16 years* \\
Commuting & Workers $\ge$ 16 years \\
Language Skills  & Population $\ge$ 5 years \\
Residence Continuity & Total Population \\
Disability Status & Civilian Noninstitutionalized Population \\
Preventable Illness (Incidence per 100,000) &  Incidence per 100,000 \\
Overcrowding & All Households \\
\hline
\end{tabular}
\end{center}
\label{tab:refPop}
\end{table}%

\begin{table}[htp]
\caption{Feature Selection for Linear Regression Models.}
\begin{center}
\begin{tabular}{|l|l|l|l|l|}
\hline
 &  Model 1 & Model 2 & Model 3 & Model 4 \\
\hline
Target Feature &
Cases per 100,000 &
Cases per 100,000 &
Deaths per 100,000 & 
Deaths per 100,000 \\
Input Features & 
Top 5 Correlations &
Top 20 Correlations & 
Top 5 Correlations&
Top 20 Correlations\\
&to Target  &  to Target; eliminate  & to Target  &  to Target; eliminate \\
& & co-correlations & &  co-correlations \\
\hline
\end{tabular}
\end{center}
\label{tab:models}
\end{table}%

\begin{table}[htp]
\caption{Goodness of Fit Test Results for Model 1}
\begin{center}
\begin{tabular}{|l|l|l|l|l|}
\hline
Model 1:  &   \multicolumn{2}{c|}{Cases}  &  \multicolumn{2}{c|}{Deaths}  \\
 & Statistic & p-Value & Statistic & p-Value  \\
 \hline
Jarque-Bera Test & 1.510 & 0.470 & 1.021 & 0.600\\
Breusch-Pagan test & 8.718 & 0.114 &  7.069 & 0.216\\
 Linear Rainbow & 1.050 & 0.474 & 1.385 & 0.269 \\
VIF values & & &  &\\
constant & 665.682 & & 568.208 &\\
Age 0-19 & 4.104& & 3.604 & \\
Hispanic Ethnicity (\%) & 7.595 & & 9.568 &  \\
Long Term Diabetes Complications (incidence per 100K) & 3.880 & & 2.414 & \\
Education: 9th to 12th grade, no diploma (\%) & 3.492 & & 2.414  &\\
Average family size & 5.270 & & 4.755 & \\
\hline
\end{tabular}
\end{center}
\label{tab:featuresModel1}
\end{table}%

\begin{table}[htp]
\caption{Goodness of Fit Test Results for Model 2}
\begin{center}
\begin{tabular}{|l|l|l|l|l|}
\hline
Model 2:  &   \multicolumn{2}{c|}{Cases}  &  \multicolumn{2}{c|}{Deaths}  \\
 & Statistic & p-Value & Statistic & p-Value  \\
 \hline
Jarque-Bera Test & 2.088 & 0.352 & 137.839 & 0.000\\
Breusch-Pagan test & 5.116 & 0.276 &  3.549 & 0.471\\
 Linear Rainbow & 0.957 & 0.547 &3.310 & 0.011\\
VIF values & & &  &\\
constant & 40.823& & 40.813 &\\
Age 0-19 &  1.267 & & 1.267 & \\
Hispanic Ethnicity (\%) & 1.316 & & 1.316 &  \\
Long Term Diabetes Complications (incidence per 100K) & 1.334 & & 1.334  & \\
Education: 9th to 12th grade, no diploma (\%) & 1.414 & & 1.414  &\\
Average family size & 5.270 & & 4.755 & \\
\hline
\end{tabular}
\end{center}
\label{tab:featuresModel2}
\end{table}%

%\section*{Acknowledgements}

% The \nocite command causes all entries in a bibliography to be printed out whether or not they are actually referenced in the text. 
%\nocite{*}

\bibliography{GoodOHare}

\end{document}